\documentclass{article}

\usepackage{microtype}
\usepackage{graphicx}
\usepackage{subfigure}
\usepackage{booktabs} 
\usepackage{multirow} 
\usepackage{hyperref}

\usepackage[accepted]{icml2025}

\usepackage{amsmath}
\usepackage{amssymb}
\usepackage{mathtools}
\usepackage{amsthm}

\usepackage[capitalize,noabbrev]{cleveref}

\theoremstyle{plain}

\theoremstyle{definition}

\theoremstyle{remark}

\usepackage[textsize=tiny]{todonotes}

\icmltitlerunning{Adapting Text-to-Speech to Environmental Contexts with Flow Matching}

\newcommand\ourmethod{\emph{UmbraTTS}}

\newcommand{\ourSNR}{SER~}
\newcommand{\ourSNREm}{\text{SEREmbed}~}

\begin{document}

\twocolumn[
\icmltitle{UmbraTTS: Adapting Text-to-Speech to Environmental Contexts with Flow Matching}



\icmlsetsymbol{equal}{*}
\icmlsetsymbol{workdone}{\dag}

\begin{icmlauthorlist}
\icmlauthor{Neta Glazer}{aiola}
\icmlauthor{Aviv Navon}{aiola}
\icmlauthor{Yael Segal}{aiola}
\icmlauthor{Aviv Shamsian}{aiola}
\icmlauthor{Hilit Segev}{aiola}
\icmlauthor{Asaf Buchnick}{aiola}
\icmlauthor{Menachem Pirchi}{ind,workdone}
\icmlauthor{Gil Hetz}{aiola}
\icmlauthor{Joseph Keshet}{aiola}
\end{icmlauthorlist}

\icmlaffiliation{aiola}{aiOla Research}
\icmlaffiliation{ind}{Independent Researcher}

\icmlcorrespondingauthor{Neta Glazer}{neta.glazer@aiola.com}
\icmlcorrespondingauthor{Aviv Navon}{aviv@aiola.com}

\icmlkeywords{TTS, Environmental Sounds}

\vskip 0.3in
]



\printAffiliationsAndNotice{\textsuperscript{\dag}Work done while at aiOla Research}  

\begin{abstract}
Recent advances in Text-to-Speech (TTS) have enabled highly natural speech synthesis, yet integrating speech with complex background environments remains challenging. We introduce UmbraTTS, a flow-matching based TTS model that jointly generates both speech and environmental audio, conditioned on text and acoustic context. Our model allows fine-grained control over background volume and produces diverse, coherent, and context-aware audio scenes. A key challenge is the lack of data with speech and background audio aligned in natural context. To overcome the lack of paired training data, we propose a self-supervised framework that extracts speech, background audio, and transcripts from unannotated recordings. Extensive evaluations demonstrate that \ourmethod{} significantly outperformed existing baselines, producing natural, high-quality, environmentally aware audios. Audio samples are available at: \textcolor{magenta}{\url{https://aiola-lab.github.io/umbra-tts/}}
\end{abstract}
\vspace{-0.85cm}
\section{Introduction}
In recent years, text-to-speech (TTS) systems have seen significant advancements, enabling the generation of highly natural and intelligible speech across a range of applications and interactive scenarios \cite{tan2021survey, zhang2023survey, casanova2022yourtts, tan2024naturalspeech, ren2020fastspeech, paul2020enhancing}. However, for TTS to be truly natural and contextually appropriate in real-world scenarios, it must not only generate clean speech, but also address how speech interacts with diverse acoustic environments and background sounds \cite{lee2024voiceldm}. In parallel, significant progress in text-to-audio (TTA) models has shown the potential of generating rich, diverse audio scenes from text~\cite{liu2023audioldm, borsos2023audiolm, vyas2023audiobox, majumder2024tango}. While these models demonstrate impressive results in general audio synthesis, they struggle in generating intelligible and coherent speech, often producing artifacts or unnatural prosody. 
As a result, the field of speech generation with diverse environmental conditions remains largely under-explored, with only a handful of works explicitly addressing this challenge~\cite{lee2024voiceldm, jung2024voicedit, kim2024speak}.

\begin{figure}[t]
\centering
\includegraphics[width=0.7\linewidth]{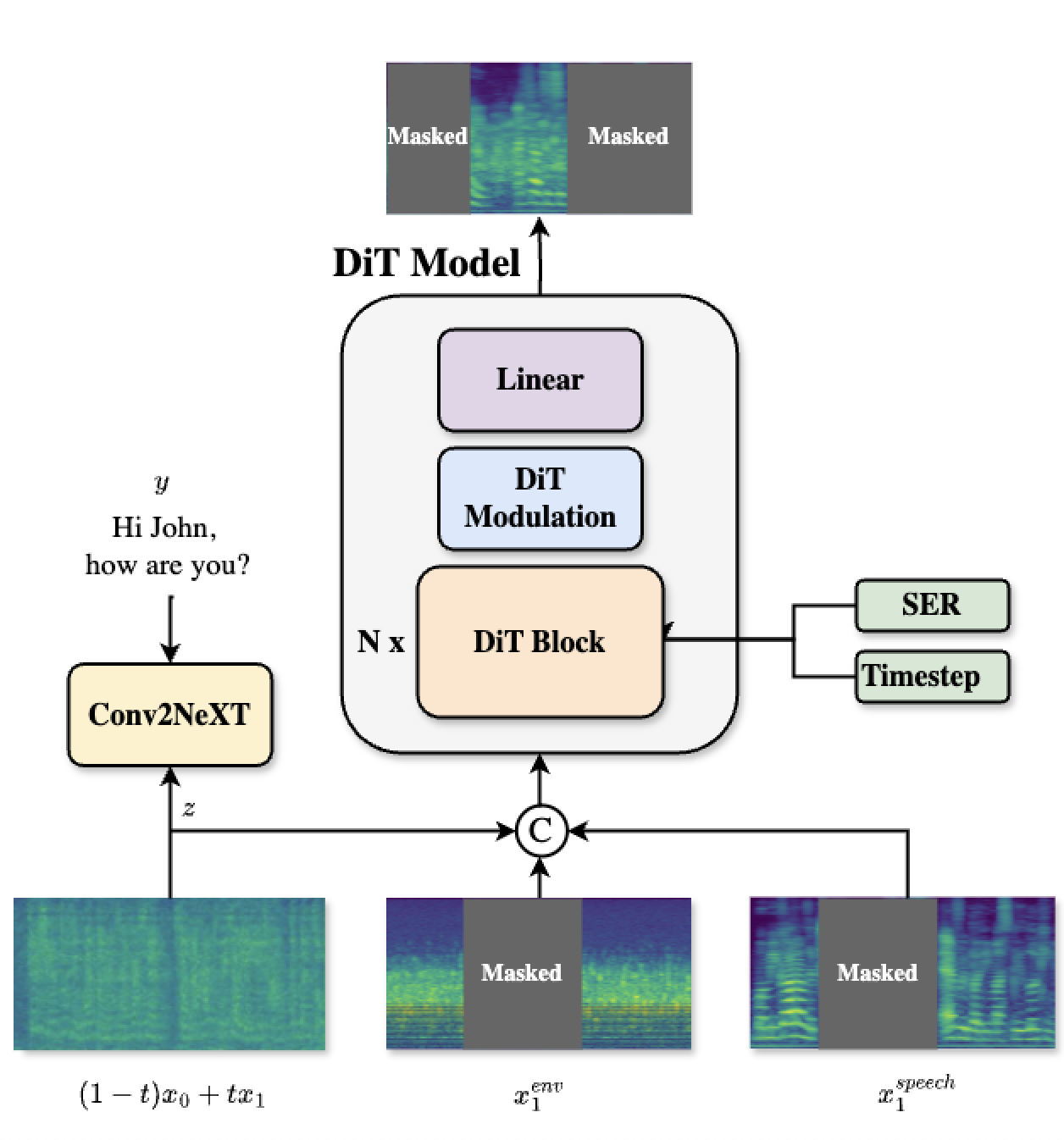}
\vspace{-0.53cm}
\caption{The \ourmethod{} architecture.  }
\label{fig:arch}
\vspace{-0.73cm}
\end{figure}

A naive approach to integrating speech with background noise is to first generate clean speech and then overlay environmental sounds. However, this method has three major limitations. First, it overlooks how speakers naturally adjust vocal effort, tone, and rhythm according to their surroundings - a behavior known as the Lombard effect \cite{zollinger2011lombard}. 
Second, it lacks coherent integration of background; for example, synchronizing applause with speech requires precise timing and context awareness, not merely mixing sounds. Third, it cannot produce varied background sound within the same condition, limiting realism and flexibility. Experiments detailed in Sec. 4.2 reveal the limitations of the naive approach, highlighting its inability to deliver natural and contextually appropriate audio.


In this work, we propose UmbraTTS, an environmentally aware TTS system that synthesizes speech alongside realistic, contextually integrated background sound.  Our method allows fine-grained control over background volume and supports diverse acoustic scenes. A key challenge in training environmentally-aware TTS models is the lack of recordings where speech and background audio co-occur naturally and are contextually aligned. To address the lack of paired data, we develop a self-supervised framework that learns from naturally mixed audio recordings. Our model and training dataset will be released to support future research in environmentally aware speech synthesis. Extensive evaluations show that UmbraTTS outperforms prior methods in both intelligibility and environmental coherence.

\begin{figure*}[t] 
    \centering    
    \begin{minipage}{0.25\textwidth}
        \centering
        \ourSNR Level: 0 \\ 
        \includegraphics[width=\textwidth]{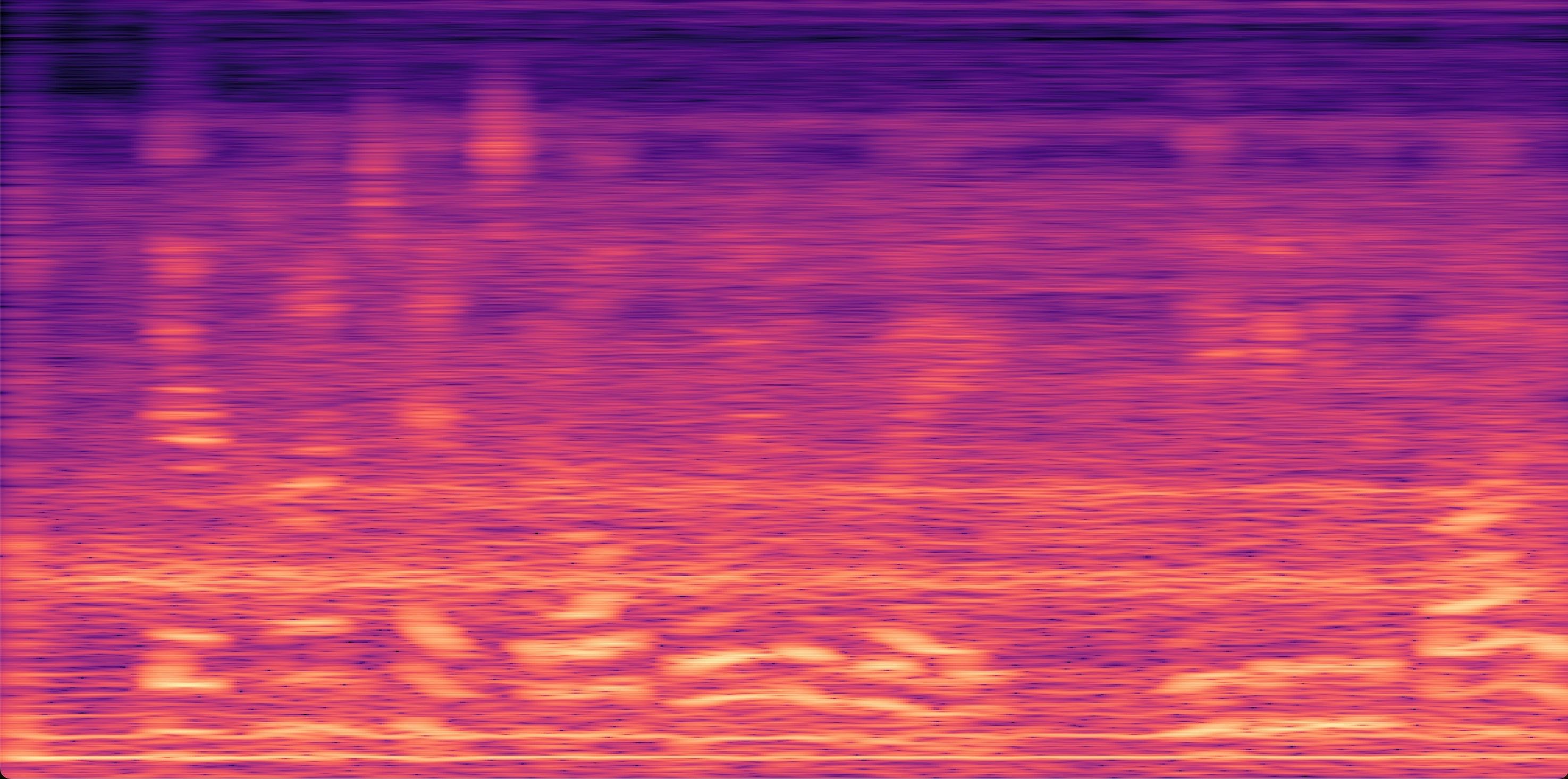}
    \end{minipage}
    \hspace{0.5cm}
    \begin{minipage}{0.25\textwidth}
        \centering
        \ourSNR Level: 0.5 \\ 
        \includegraphics[width=\textwidth]{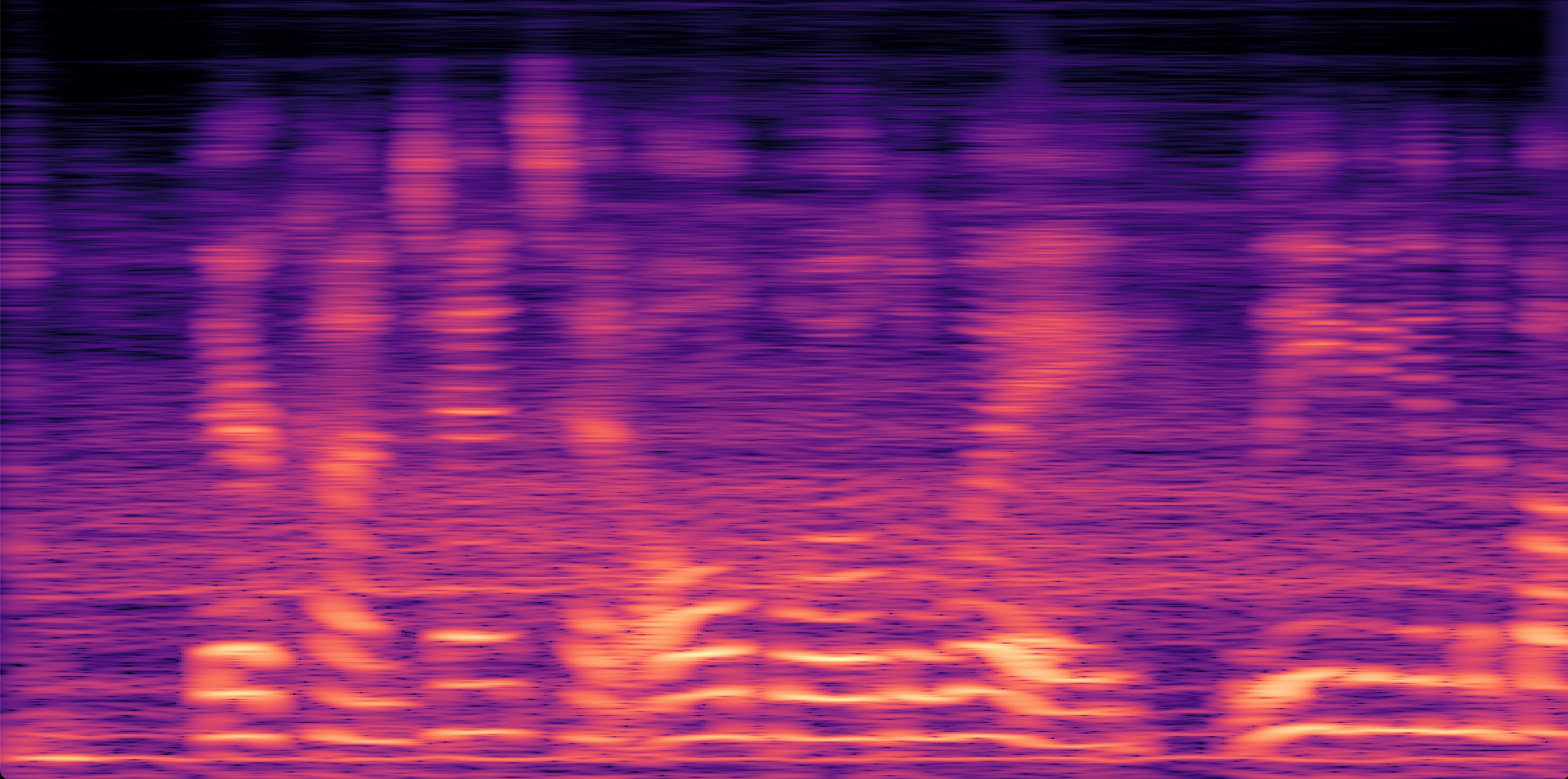}
    \end{minipage}
    \hspace{0.5cm}
    \begin{minipage}{0.25\textwidth}
        \centering
        \ourSNR Level: 1.0 \\ 
        \includegraphics[width=\textwidth]{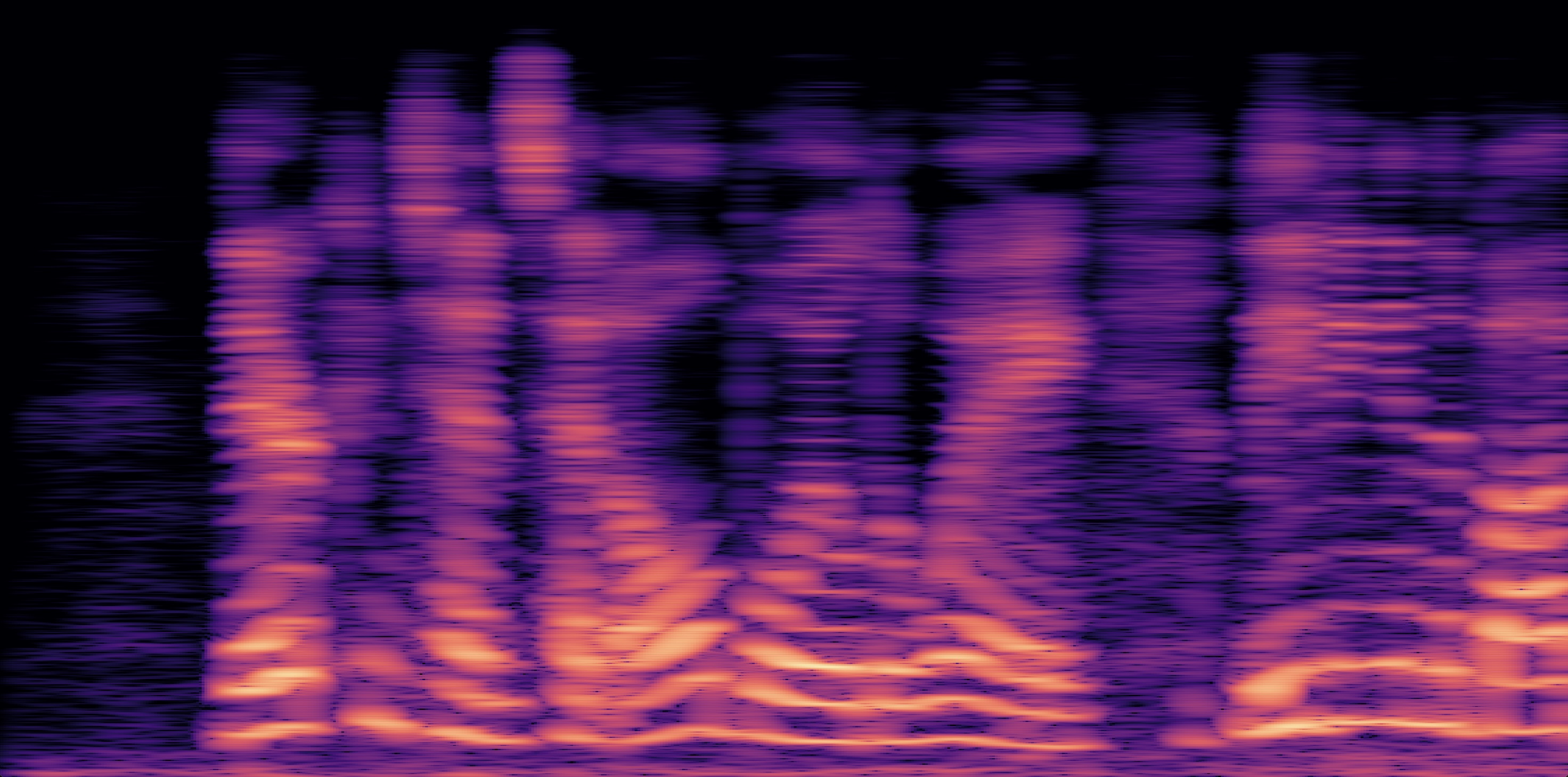}
    \end{minipage}
    \vspace{-0.25cm}
    \caption{Generated Mel spectrograms under different \ourSNR conditioning levels, synthesized with the same text conditioning.}
    \label{fig:mel_spectrograms}
    \vspace{-0.4cm}
\end{figure*}

\vspace{-0.25cm}
\section{Related Works}
Recently, audio generation has advanced with models such as AudioLDM \cite{liu2023audioldm}, AudioLDM2 \cite{liu2023audioldm}, AudioGen \cite{kreuk2022audiogen}, Make-An-Audio \cite{huang2023make} and Fugatto \cite{vallefugatto}. These models can create high-quality audio and soundscapes from text or other inputs.
A specific subset of these models is environmentally-aware TTS models, which have mostly been explored through diffusion-based approaches, such as VoiceLDM \cite{lee2024voiceldm},  VoiceDiT \cite{jung2024voicedit} and Speak-in-the-Scene framework \cite{kim2024speak}.
Flow matching architectures \cite{lipman2022flow} have demonstrated superior synthesis quality in generative models, specifically for speech synthesis tasks~\cite{chen2024f5,navon2025flowtse}. Models like E2-TTS \cite{eskimez2024e2}, 
MatCha-TTS \cite{mehta2024matcha}, and Voiceox \cite{le2024voicebox} use direct probability paths for high-fidelity generation, but with no explicit environmental conditioning.

\vspace{-0.25cm}
\section{Proposed Method}

We introduce \ourmethod{}, a flow-matching \cite{lipman2022flow} based method for environmental-aware text-to-speech. Following the F5-TTS framework \cite{chen2024f5}, which demonstrated the effectiveness of flow matching for text-to-speech, we propose an advanced architecture for environmental conditioning. The architecture of our method is illustrated in Fig.~\ref{fig:arch} and described below.
\vspace{-0.3cm}
\paragraph{Flow Matching.} The flow matching objective is to match a time-dependent probability path $p_t \, {\scriptstyle (0 \leq t \leq 1)}$ from a simple source distribution $p_0 = p$, e.g., the standard normal distribution $\mathcal{N} (0, I)$, 
to $p_1$ which approximates the data distribution $q$. 
The transformation of the probability path is modeled by a neural network \(v_t\) parameterized by $\theta$, which is trained to estimate a time-dependent vector field \( u_t: [0,1] \times \mathbb{R}^d \to \mathbb{R}^d \). From this vector field, we define a transformation 
 \( \psi_t: [0,1] \times \mathbb{R}^d \to \mathbb{R}^d \) satisfying the ordinary differential equation (ODE);
\(\frac{d}{dt} \psi_t(x) = u_t(\psi_t(x))\) 
where \( \psi_t(x) \) represents the flow transitions from $p_0$ to $p_1$. However, directly constructing \( p_t \) and its generating field \( u_t \) is intractable in practice. To address this, training is performed using conditional probability paths. The resulting Conditional Flow Matching (CFM) loss has been shown to yield gradients equivalent to those of the original FM objective~\cite{lipman2022flow}, and is defined as:
$$\mathcal{L}_{\text{CFM}}(\theta) = \mathbb{E}_{t, x_t, x_1} \lVert v_t ^\theta(x_t ) - u_t(x_t | x_1) \rVert^2,$$
where \( x_1 \sim q \) is a sample from the data distribution, and \( x_t = \psi_t(x_0) \) denotes the flow at time \( t \) starting from \( x_0 \sim p \). When using the Optimal Transport (OT) displacement form \( \psi_t(x) = (1 - t)x + t x_1 \), the CFM loss simplifies to:
$$\mathcal{L}_{\text{CFM}}(\theta) = \mathbb{E}_{t, x_0, x_1} \lVert v_t^\theta(x_t) - (x_1 - x_0) \rVert^2,$$
For the neural network \(v_t\), we utilize Diffusion Transformer (DiT) blocks and adaptive LayerNorm (adaLN-zero) to process flow time step conditioning.

\vspace{-0.3cm}
\paragraph{Training Methodology.}
\label{sec:Training-Methodology}
We train our model to reconstruct a missing audio segment, which includes speech and environmental background, using its surrounding speech and environmental audio separately, along with the complete text transcription. The training data consists of audio utterances including speech and environmental sounds, paired with their corresponding transcripts. During training, the model receives acoustic inputs in the form of mel spectrogram features, represented as $x \in \mathbb{R}^{F \times N}$, 
where $F$ denotes the mel dimension and $N$ is the sequence length.
We denote the target audio mel spectrogram as $x_1$, the target audio's transcript as $y$, and the audio components' mel spectrograms as $x_1^{speech}$, $x_1^{env}$.
Within the CFM framework, the model receives the noisy mel representation $(1 - t)x_0 + t x_1$ alongside the masked target mel spectrogram $(1 - m) \odot x_1$, where $x_0$ is sampled Gaussian noise, $t$ is a randomly sampled flow step, and $m \in \{0, 1\}^{F \times N}$ is a binary temporal mask. The target text $y$ of $x_1^{speech}$, is tokenized into a characters sequence and padded with filler tokens to match the length of $x_1$, forming the extended sequence $$
    z = g(c_1, c_2, \dots, c_S, \underbrace{\text{F}, \dots, \text{F}}_{(N-S) \text{ elements}})$$
where, $c_i$ represents the $i$-th character in the sequence, and $\text{F}$ denotes the filler token. The function $g$ is an embedding network \emph{ConvNeXT V2} \cite{woo2023convnext}.
We then train our model to reconstruct $m\odot x_1$, using $(1 - m) \odot x_1^{\text{speech}}$, $\quad (1 - m) \odot x_1^{\text{env}}$, and $z$. 
At inference, samples from \( p_1 \) are obtained by reversing the noisy trajectory, integrating the learned velocity field from noise \( x_0 \) to clean speech \( x_1 \).

\vspace{-0.3cm}
\paragraph{Speech-to-Environment Ratio.}\label{sec:ser_in_training}



For real-world applications, the ability to control background volume improves diversity, naturalness, and intelligibility in speech synthesis. We introduce speech to environment ratio as SER $\in[0,1]$, representing the ratio of the speech compared to the environment, where lower SER values indicate high environmental sound level relative to speech. We train our model to allow controlling SER level at inference time. Practically, SER values are inserted to the SEREmbed, consist of a sinusoidal positional encoding followed by a multi-layer perceptron (MLP). The \ourSNREm output is combined with the flow step embedding \text{TimeEmbed}: $ c = \text{TimeEmbed}(t) + SEREmbed(\textit{$ser$}),$ where $t$ represents the flow time step. Then, as done in previous works \cite{peebles2023scalable}, we incorporate the SER conditioning through $c$, by regressing the adaLN-Zero block parameters.
\vspace{-0.3cm}
\paragraph{Inference.}
To generate speech with the desired content, we condition the model on the mel spectrogram of a reference speech \( x^{\text{speech}} \) its transcription \( y^{\text{speech}}\), a target environment prompt \( x^{\text{env}} \), and a target text prompt \( y^{\text{gen}} \). The reference speech, along with its transcript, provides speaker characteristics, the reference environmental audio provides environmental characteristics, and the text prompt \(y^{\text{gen}} \) guides the content. To estimate the duration of the generated speech, we use the ratio of character lengths between \( y^{\text{gen}} \) and \( y^{\text{speech}} \), assuming it does not exceed the mel spectrogram length. Padding with filler tokens is applied as in training.
At inference, we sample \( x_0 \sim p_0 \) and use an ODE solver to integrate the learned velocity field from \( t=0 \) to \( t=1 \). The reference portion of the generated mel spectrogram is discarded, and a vocoder converts the remaining mel spectrogram into a waveform.
\vspace{-0.3cm}
\paragraph{Self-Supervised Data Generation.}
\label{sec:ssl} A core challenge in training environmentally aware TTS models is the lack of datasets with isolated speech, background sounds, and aligned transcripts. Existing data typically contains either clean speech or mixed audio without annotations \cite{he2024emilia}. To address this, we introduce a self-supervised approach using unlabeled audio recordings with natural speech-background mixtures. From each recording $x_1$, we construct a triplet: $x^{speech}_1$ (clean speech), $x^{env}_1$ (background audio), and $y$ (transcript). We transcribe $x_1$ using Whisper-large-v2~\cite{radford2023robust} and then apply one of two separation strategies: (i) Voice Activity Detection (VAD)~\cite{SileroVAD} is used to detect non-speech regions in \( x_1 \); these segments are extracted and concatenated to form \( x^{env}_1 \), which represents the environmental sound. This method is effective when speech and background sounds are clearly separable in time.
(ii) Source separation~\cite{zhao2024mossformer2} handles overlapping sources by splitting \( x_1 \) into \( x^{speech}_1 \) and \( x^{env}_1 \).
During training, one of the two methods is randomly selected, improving the robustness to varying audio complexities and overlap levels.
\vspace{-0.65cm}
\section{Experiments}
\textbf{Datasets.} We trained our model as outlined in Section \ref{sec:Training-Methodology}, using two datasets; (i) AudioSet \cite{gemmeke2017audio}, where each audio sample naturally contains both environmental sounds and speech, and (ii) LibriSpeech \cite{panayotov2015librispeech}, consists of clean speech recordings. To add environmental conditions with a specific \ourSNR value, we combined LibriSpeech recordings with environmental sounds from the FSD50K dataset \cite{fonseca2021fsd50k}, creating target samples with both speech and background noise at controlled SNR levels. 
We randomly sampled SNR values between -5 and 20 dB and normalized the values to be in the range of 0 to 1 to create \ourSNR values. Then, we scaled and added environment audio to waveform per SER.
\vspace{-0.2cm}

\textbf{Training setup.} We trained our model with 550k training steps on two L40S GPUs, 11.5k audio frames at each batch. We used AdamW optimizer, with a learning rate of 5e-5.
\vspace{-0.2cm}

\textbf{Baselines.} For the TTS task, we compare our model against several SOTA baselines; VoiceLDM \cite{lee2024voiceldm}, which is a TTS model conditioned on both textual input and an audio description of the environment. VoiceDiT \cite{jung2024voicedit}, a TTS model conditioned on textual environmental description. WavCraft \cite{liang2024wavcraft}, an LLM-based approach that integrates target background audio with clean TTS-generated speech samples. Notably, WavCraft does not synthesize environmental sounds but instead blends them with the generated speech.
For the audio generation tasks, we extend our baseline comparisons by incorporating text-to-audio model: AudioLDM \cite{liu2023audioldm} and AudioLDM2 \cite{liu2024audioldm}.

\vspace{-0.25cm}
\subsection{Quantitative Evaluation}
First, we evaluated our model’s capability to generate speech with integrated environmental sounds. To assess this, we use the AudioCaps test subset as described in VoiceLDM \cite{lee2024voiceldm}, which contains audio recordings with both speech and background noise. During evaluation, our model is conditioned on the target text and the background sound from the test subset. The results are presented in Table \ref{tab:audiocaps_tts}. For quantitative evaluation, we report the Frechet Audio Distance (FAD) \cite{kilgour2018fr}, Kullback-Leiber (KL), word error rate (WER), computed using Whisper large-v3 \cite{radford2023robust}, and the CLAP score \cite{wu2023large}, measured between the caption and the generated audio. The results show that \ourmethod{} achieves SOTA performance compared to commonly used baselines.
Next, we evaluated our model’s audio-to-audio capabilities. In this experiment, we conditioned our model only on the environmental target audio. The results are present in Table \ref{tab:audiocaps_ata} for both the AudioCaps and MusicCaps. Again, our method achieves SOTA results across all evaluation metrics except for the FAD on AudioCaps, where it remains competitive.

\begin{table}
\caption{Quantitative evaluation on AudioCaps.}
\centering
\scriptsize
\setlength{\tabcolsep}{3pt}
\renewcommand{\arraystretch}{1.0}
\begin{tabular*}{\linewidth}{@{\extracolsep{\fill}}lccc}
    \toprule
    Method & WER (\%) $\downarrow$ & CLAP $\uparrow$ & FAD $\downarrow$ \\
    \midrule
    Ground Truth & $17.47$ & $0.40$ & - \\
    VoiceLDM\textsubscript{\scriptsize{audio}} & $13.46$ & $0.21$ & $7.03$ \\
    VoiceLDM\textsubscript{\scriptsize{text}} & $10.39$ & $0.21$ & $5.56$ \\
    VoiceDiT & $7.09$ & $0.22$ & $4.60$ \\
    WavCraft & $21.16$ & \textbf{0.40} & $8.16$ \\
    \midrule
    \ourmethod{} & \textbf{6.89} & $0.37$ & \textbf{4.14} \\
    \bottomrule
\end{tabular*}
\vspace{-0.4cm}
\label{tab:audiocaps_tts}
\end{table}

\begin{table}[t]
\centering
\caption{Quantitative evaluation on AudioCaps and MusicCaps.}
\scriptsize
\setlength{\tabcolsep}{3pt}
\renewcommand{\arraystretch}{1.0}
\begin{tabular*}{\linewidth}{@{\extracolsep{\fill}}lcccccc}
\toprule
& \multicolumn{3}{c}{AudioCaps} & \multicolumn{3}{c}{MusicCaps} \\
\cmidrule(lr){2-4} \cmidrule(lr){5-7}
& CLAP $\uparrow$ & KL $\downarrow$ & FAD $\downarrow$ & CLAP $\uparrow$ & KL $\downarrow$ & FAD $\downarrow$ \\
\midrule
VoiceLDM\textsubscript{\scriptsize{audio}} & $0.51$ & $10.02$ & $4.13$ & $0.63$ & $4.64$ & $5.72$ \\
VoiceLDM\textsubscript{\scriptsize{text}} & $0.29$ & $2.95$ & $10.91$ & $0.47$ & $1.47$ & $3.61$ \\
VoiceDiT & $0.45$ & $1.87$ & $3.55$ & -- & -- & -- \\
AudioLDM & $0.42$ & $2.01$ & $4.27$ & $0.54$ & $1.42$ & $4.15$ \\
AudioLDM2 & $0.58$ & $2.36$ & $\textbf{2.04}$ & $0.61$ & $3.83$ & $3.51$ \\
\midrule
\ourmethod{} & $\textbf{0.619}$ & $\textbf{1.87}$ & $2.65$ & $\textbf{0.77}$ & $\textbf{0.939}$ & $\textbf{3.11}$ \\
\bottomrule
\end{tabular*}
\vspace{-0.55cm}
\label{tab:audiocaps_ata}
\end{table}

\begin{figure}[t]
\centering
\includegraphics[width=0.85\linewidth]{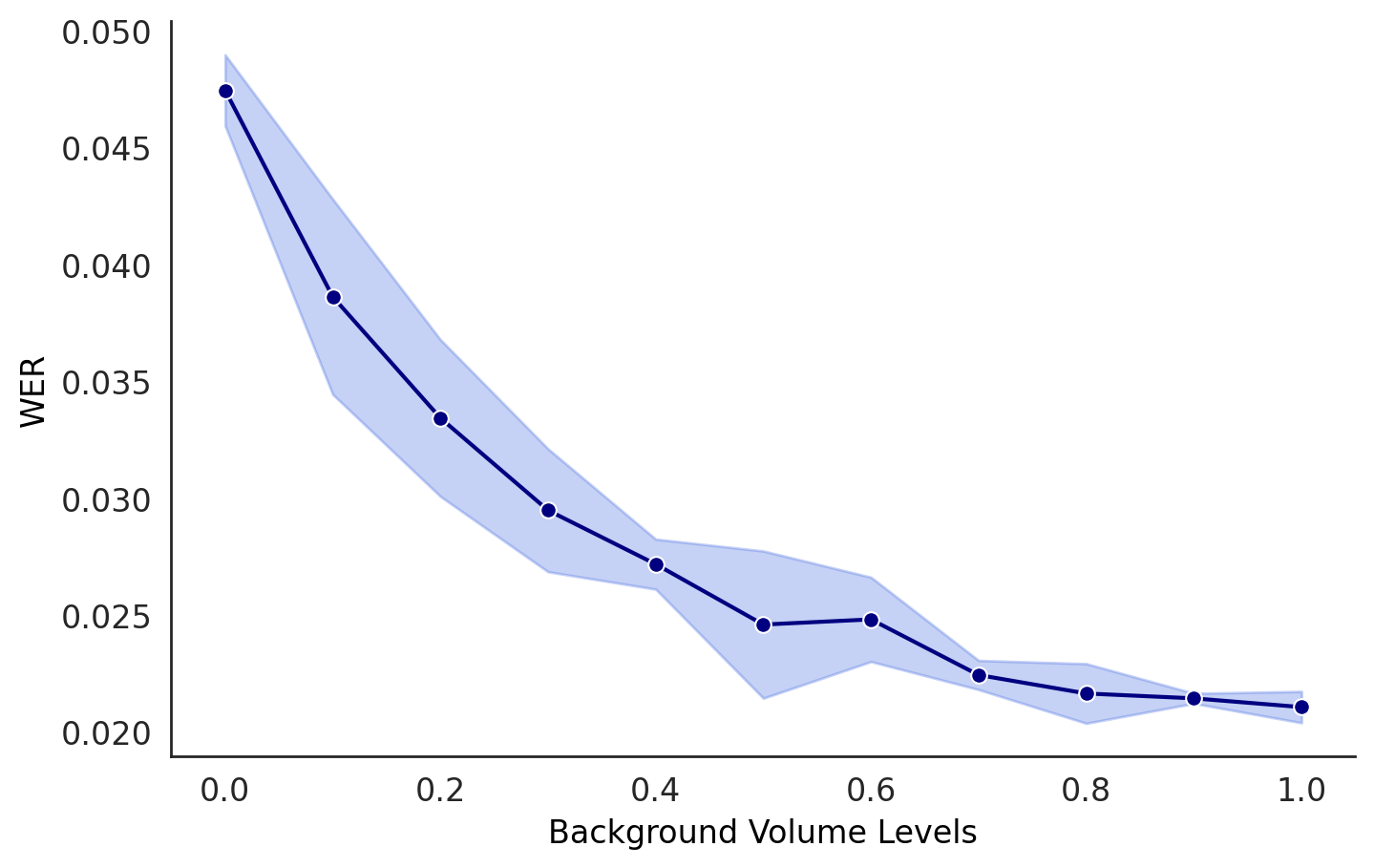}
\vspace{-0.4cm}
\caption{\textit{SER vs WER}}
\label{fig:SER-vs-WER}
\vspace{-0.7cm}
\end{figure}
\vspace{-0.25cm}
\subsection{Qualitative Human Evaluation}
\label{sec:Qualitative_eval}
\textbf{Environmental Integration.} We conducted an A/B test to compare our method with the natural and naive baseline in which background audio is simply added to clean speech. Specifically, we use F5-TTS \cite{chen2024f5} to synthesize speech from text and then mix it with real background recordings. 
We randomly select 30 samples from the AudioSet test set and generate paired outputs: one using our method, and the other using F5-TTS with background sounds added post hoc. 
Participants were asked: \emph{In which audio is the speech more naturally and clearly integrated with the background noise?} Each pair was evaluated by at least 20 unique raters. Our results show that \textbf{87.89\%} of responses preferred \ourmethod{}. This highlights the importance of generating speech and background jointly rather than combining them post hoc. 

Next, we conduct an A/B listening test comparing \ourmethod~with the state-of-the-art baseline, VoiceLDM. We randomly select 55 audio samples from the AudioCaps test set for this test. Each participant hears the real audio and two generated versions—UmbraTTS and VoiceLDM. Participants then answered three questions for each set: (i) Which generated version sounds more natural? (ii) Which version better preserves the original background audio? (iii) Which version integrates the background audio better with the speech content?. The evaluation included 50 participants across diverse ages, genders, and backgrounds, with each question answered by at least 20 unique individuals.
The results, summarized in Table \ref{tab:ab_test_ttsa}, reveal that \ourmethod~ outperformed VoiceLDM across all metrics. Each percentage reflects the proportion of total A/B test responses in which our method was preferred, 
indicating a clear listener preference and underscoring its effectiveness.

\textbf{Controlled SER.} To evaluate the effectiveness of our model in controlling the speech-to-environment ratio (SER), we conducted a controlled listening test with 25 participants. We generate 20 pairs of audio samples, where each pair contains audio with identical speech content and environmental conditions but differed only in their \ourSNR values, sample uniformly from the range [0, 1]. Participants were presented with each pair and asked to identify the audio sample in which the environmental sound appeared louder relative to the speech. The responses of the participants aligned with the intended control values of our model \ourSNR \textbf{96. 6\%} of the time, confirming that our method successfully achieves precise and noticeable control over the relative volume of environmental sounds. Also, we analyzed the effect of different \ourSNR settings on speech intelligibility by plotting the Word Error Rate (WER) against varying \ourSNR levels. The results, averaged over three random repetitions, are depicted in Fig. \ref{fig:SER-vs-WER}. Furthermore, Fig. \ref{fig:mel_spectrograms} illustrates how altering \ourSNR values impacts the Mel spectrogram, visually demonstrating clear changes in the prominence of environmental sounds relative to speech.
\begin{table}
\caption{A/B test human evaluation for \ourmethod{} vs. VoiceLDM.}
\centering
\scriptsize
\setlength{\tabcolsep}{3pt}
\renewcommand{\arraystretch}{1.1}
\begin{tabular*}{\linewidth}{@{\extracolsep{\fill}}l c}
    \toprule
    Evaluation Metric & A/B Test Preference (\%) $\uparrow$ \\
    \midrule
    Naturalness & 81.91 \\
    Background Integration & 78.54 \\
    Background Preservation & 81.83 \\
    \bottomrule
\end{tabular*}
\vspace{-0.76cm}
\label{tab:ab_test_ttsa}
\end{table}



\vspace{-0.3cm}
\section{Discussion}
\vspace{-0.15cm}
In this work, we introduce \ourmethod{}, a novel environmentally aware TTS model based on conditional flow matching. To overcome the challenge of limited paired data, we also propose an efficient SSL framework for training \ourmethod{}. We evaluate our approach across diverse setups, including human evaluations, and demonstrate the superiority of \ourmethod{} over recent TTS baselines. We believe our approach paves a new path for environmentally aware TTS, enabling more natural and controllable synthesis.


\bibliography{mybib}
\bibliographystyle{icml2025}




\end{document}